\RequirePackage{fix-cm}
\documentclass{svjour3}                     
\smartqed  
\usepackage{graphicx}
\newcommand{\be}{\begin{equation}}
\newcommand{\ee}{\end{equation}}
\def\dif{\mathop{\rm \,d}\nolimits}
\def\ci{\mathop{\textrm{i}}\nolimits}

\begin{document}


\title{Birkhoff theorem and conformal Killing-Yano tensors}

\titlerunning{Birkhoff theorem}        

\author{Joan Josep Ferrando        \and
   Juan Antonio S\'aez 
}


\institute{Joan Josep Ferrando \at
              Departament d'Astronomia i Astrof\'{\i}sica, Universitat
de Val\`encia, E-46100 Burjassot, Val\`encia, Spain. \\
Observatori Astron\`omic, Universitat
de Val\`encia, E-46980 Paterna, Val\`encia, Spain. \\
                           \email{joan.ferrando@uv.es}           
           \and
          Juan Antonio S\'aez  \at
               Departament de Matem\`atiques per a l'Economia i l'Empresa,
Universitat de Val\`encia, E-46071 Val\`encia, Spain
}

\date{Received: date / Accepted: date}

\maketitle

\begin{abstract}
We analyze the main geometric conditions imposed by the hypothesis of the Jebsen-Birkhoff theorem.
We show that the result (existence of an additional Killing vector) does not necessarily require a
three-dimensional isometry group on two-dimensional orbits but only the existence of a conformal Killing-Yano tensor.
In this approach the (additional) isometry appears as the known invariant Killing vector that the ${\cal D}$-metrics admit.

\end{abstract}

\section{Introduction}

The pioneering results by Jebsen \cite{Jebsen} and Birkhoff \cite{birkhoff} are applied to spherically
symmetric vacuum solutions. Subsequently, several generalizations have been accomplished concerning
either the energy content (two double eigen-values \cite{cahen-debever}), or the isometry group
(plane and hyperbolical on space-like orbits \cite{goener}, or maximal symmetry on time-like orbits \cite{barnes}).

Thus, the two hypotheses of the original Jebsen-Birkhoff theorem have been weakened. Our aim here is to show
 that the present version of the former, the existence of a maximal group of symmetries on two-dimensional
 non-null orbits, admits a weaker statement.

In this paper we work on an oriented space-time with a metric tensor
$g$ of signature $\{-,+,+,+\}$ and metric volume element $\eta$. The Riemann, Ricci and Weyl tensors
are defined as given in \cite{kramer} and are denoted, respectively,
by $Riem$, $Ric$ and $W$. For the metric product of two vectors we
write $(X,Y) = g(X,Y)$. If $A$ and $B$
are 2-tensors, $A \cdot B$ denotes the 2-tensor $(A \cdot
B)^{\alpha}_{\ \beta} = A^{\alpha}_{\ \mu} B^{\mu}_{\ \beta}$, $A^2
= A \cdot A$, $A(X,Y) = A_{\alpha \beta} X^{\alpha} Y^{\beta}$, $A(X) = A_{\alpha \beta} X^{\beta}$, and
$(A,B) = \frac12 A_{\alpha \beta} B^{\alpha \beta}$. For a vector $X$ and a (p+1)-tensor $t$, $i(X)t$ denotes the
inner product, $[i(X)t]_{\underline{p}} = X^{\alpha}t_{\alpha
\underline{p}}$, the underline denoting multi-index. And if $\omega$ is a (p+1)-form, $\delta \omega $ denotes its exterior codiferential, $(\delta \omega)_{\underline{p}} = -\nabla_{\alpha} \omega^{\alpha}_
{\ \underline{p}}$.

Note that the existence of a maximal group of symmetries on two-dimensional non-null orbits   
implies that the metric is conformal to a 2+2 product one with a restricted
conformal factor \cite{bona}, which leads to a 2+2 warped product space-time. Then, a
non null Killing-Yano tensor $A$ exists \cite{fs2+2}. 
  
Now, we impose a weaker condition: the existence of a conformal Killing-Yano (CKY) tensor $A$. Then, the associated self-dual two-form
  ${\cal A} = \frac{1}{\sqrt{2}} (A - \ci *A)$ satisfies the CKY equation \cite{fs-KY}:
\begin{equation} \label{CKYeq}
3 \nabla {\cal A} =  2 i({\cal Z}) {\cal G} \, , \qquad  {\cal Z} \equiv \delta {\cal A}  \, ,
\end{equation}
where $*$ is the Hodge dual operator. The 4-tensor ${\cal G}$ is the endowed metric
on the 3-dimensional complex space of the self-dual two-forms, ${\cal G}=\frac{1}{2}(G-\textrm{i} \; \eta)$, $G$
being the metric on the space of 2--forms, $G_{\alpha \beta \gamma \delta} = g_{\alpha
\gamma} g_{\beta \delta} - g_{\alpha \delta} g_{\beta \gamma}$. The self-dual Weyl tensor is ${\cal W}=\frac{1}{2}(W-\textrm{i} \; * W)$.

\section{Main results}
We know (see \cite{glass-kress} and references therein) that the integrability conditions of the
CKY equation (\ref{CKYeq}) lead to constraints on the Petrov-Bel type. Indeed, if ${\cal A}$ is a non
null two-form then the space-time is type O or type D, and in type D case ${\cal A}$ is an eigen-two-form
of the Weyl tensor: ${\cal A}  = e^{\lambda}{\cal U}$, ${\cal U}$ being the simple unitary Weyl eigen-two-form.
On the other hand, if ${\cal A}$ is a null two-form then the space-time is type O or type N, and in type N
 case ${\cal A}$ is a null eigen-two-form of the Weyl tensor: the self-dual Weyl tensor writes
 ${\cal W} = {\cal H} \otimes {\cal H}$, and ${\cal A}  = e^{\mu}{\cal H}$.

Moreover, the integrability conditions of the CKY equation also constrains the Ricci tensor. More precisely, from the Ricci identities for the two-form ${\cal A}$ and taking into account (\ref{CKYeq}) we obtain:
\begin{equation} \label{CKYci}
2 {\cal L}_{\cal Z} g = 3 [ {\cal A} , Ric] \, ,
\end{equation}
where $ {\cal L}_{\cal Z}$ denotes the Lie derivative with respect the vector field ${\cal Z}$, and for two 2-tensors $P, Q$, $[P,Q]$ denotes their commutator, $[P,Q]= P\cdot Q - Q \cdot P$. The commutator $[ {\cal A} , Ric]$
 vanishes if, and only if, $[ A , Ric] =  [*A , Ric] =0$. Then, we can state the following.

\begin{theorem} \label{theo-Killing}
If a space-time admits a conformal Killing-Yano tensor ${\cal A}$, then ${\cal Z} \equiv \delta {\cal A}$
is a (complex) Killing vector (or it vanishes) if, and only if, $[ {\cal A} , Ric]=0$.\\
When ${\cal A}$ is a non null two-form this condition states that the Ricci tensor is of Segr\`e types $[(11)(11)]$ or $[(1111)]$.\\
When ${\cal A}$ is a null two-form this condition states that the Ricci tensor is of Segr\`e types $[(211)]$ or $[(1111)]$.
\end{theorem}
{\bf Remark 1.} As commented above, a CKY tensor is an eigen-two-form of the Weyl tensor. Theorem \ref{theo-Killing} states
that ${\cal Z}$ is a Killing vector if, and only if, the Ricci geometry is also aligned with ${\cal A}$. Thus when
${\cal A}  = e^{\lambda }{\cal U}$ is a non null two-form then the Ricci tensor writes
 $Ric = -\kappa \Pi + \Lambda g$, $\Pi = 2 \, {\cal U} \cdot \bar{{\cal U}}$ being the 2+2 structure tensor, and where $\kappa$ and $\Lambda$ are two Ricci-invariant scalars.   On the other hand, when ${\cal A}  = e^{\mu}{\cal H}$ is a null two-form then the Ricci tensor writes $Ric = \sigma \ell \otimes \ell + \Lambda g$, where $\ell$ is the null fundamental vector of ${\cal H}$.
\\[2mm]
{\bf Remark 2.} Tachibana \cite{tachibana} obtained a similar equation to (\ref{CKYci}) for a n-dimensional Riemann space,
and he concludes that for an Einstein space, ${\cal Z}$ is a Killing vector. Theorem \ref{theo-Killing}
 above generalizes this result for the case of a four-dimensional space-time by considering all the compatible Ricci tensors.
 Thus it gives not only a necessary condition but also a sufficient one in order for ${\cal Z}$ to be a Killing vector.
 The particular case when $A$ is a Killing-Yano tensor has previously been considered \cite{dietz-rudiger}.
\\[2mm]
{\bf Remark 3.} Hougston and Sommers \cite{hougshton-sommers} proved that the ${\cal D}$-metrics
(vacuum type D metrics and their charged counterpart) admit a complex Killing vector given by the divergence
of a non null CKY tensor. This Killing vector is invariant since it is a Weyl concomitant. Hougston and Sommers do not quote the Tachibana paper but their result applies to a family of both
 vacuum and non vacuum solutions which cover the physically significant space-times for which theorem \ref{theo-Killing} applies.
 For short, in what follows we denote $\tilde{\cal D}$-metrics the non conformaly flat space-times admitting a non null CKY
  tensor ${\cal A}$ such that $[ {\cal A} , Ric]=0$.  Elsewhere \cite{fsEM-sym} we have shown that the ${\cal D}$-metrics are
  the $\tilde{\cal D}$-metrics with $\Lambda= constant$, and we have extended some known properties of the ${\cal D}$-metrics
  to the $\tilde{\cal D}$-metrics.

The CKY equation (\ref{CKYeq}) for the non null two-form ${\cal A}  = e^{\lambda }{\cal U}$ is equivalent to
the umbilical and Maxwellian character of the 2+2 structure defined by ${\cal U}$. These two properties mean that its principal directions are geodesic shear-free congruences and they determine a non null solution of the source-free Maxwell equations, and they  can be written, respectively,
in terms of $\{\lambda, \ {\cal U}\}$ as \cite{fs-KY}:
\begin{equation} \label{um-max}
\nabla {\cal U} = i(\delta {\cal U})[{\cal U} \otimes {\cal U}+{\cal G}] \, , \qquad
\, {\cal U}(\delta {\cal U}) =  \mbox{d} \lambda \, .
\end{equation}
From this latter equation we obtain the following expression for the Killing vector:
\begin{equation} \label{Z}
{\cal Z} \equiv \delta {\cal A} = \frac32 e^{\lambda} \delta {\cal U}  \, .
\end{equation}

In what follows we restrict ourselves to the space-times with non constant curvature by considering either
 $\tilde{\cal D}$-metrics or conformally flat metrics with a Ricci tensor $Ric = -\kappa \Pi + \Lambda g$, $\kappa \not= 0$.
  Under this assumption ${\cal U}$ is a Riemann invariant two-form and then (\ref{um-max}) and (\ref{Z}) imply:
\begin{equation} \label{invariances}
   {\cal L}_{\cal Z}{\cal U} = 0 \, , \qquad    {\cal L}_{\cal Z} \bar{\lambda} = 0\, , \qquad {\cal L}_{\cal Z} \bar{\cal U}=0 \, . 
\end{equation}
From these constraints (see \cite{fs-buit} for a similar reasoning) we obtain the following expression for the Killing two-form associated with ${\cal Z}$:
\begin{equation} \label{dZ}
\dif {\cal Z} = a \,  {\cal U} + m  \, \bar{{\cal U}} + \frac43 e^{-\bar{\lambda}}
 \, \bar{{\cal G}}({\cal Z} \wedge \bar{\cal Z}) \cdot
\bar{{\cal U}}   \, ,
\end{equation}
where for a double two-form $W$ and a two-form $F$, $W(F)$ denotes the two-form
$W(F)_{\alpha \beta} = \frac12 W_{\alpha \beta \mu \nu} F^{\mu \nu}$. Then, from (\ref{Z}), (\ref{invariances}) and (\ref{dZ}) we can easily show the following.

\begin{proposition} \label{prop-inv}
If a non constant curvature space-time admits a non null conformal Killing-Yano tensor $A$ such that $[A , Ric]=[*A, Ric]=0$,
then $Z_1 \equiv \delta A$ and $Z_2 \equiv \delta \! * \! A $ are Killing vectors (or they vanish) verifying:
(i) The CKY tensor is $Z_i$-invariant: ${\cal L}_{Z_1} A = {\cal L}_{Z_2} A = 0$,
(ii) If $Z_1 \wedge Z_2 \not=0$, they define a commutative algebra: $[Z_1, Z_2] = 0$.
\end{proposition}

Condition $Z_1 \wedge Z_2 =0$ characterizes the Kerr-NUT solutions in the set of the ${\cal D}$-metrics \cite{hougshton-sommers}  \cite{fsEM-sym}.
Their properties can be extended to the $\tilde{\cal D}$-metrics and to the conformally flat case. Indeed,
from (\ref{um-max}), (\ref{Z}) and (\ref{dZ}) (see \cite{fsEM-sym} and \cite{fs-buit} for a similar reasoning) we obtain the following.

\begin{proposition} \label{prop-KNUT}
In a non constant curvature space-time admitting a non null conformal Killing-Yano tensor $A$ such
that $[A , Ric]=[*A, Ric]=0$, let us consider $Z_1 \equiv \delta A$ and $Z_2 \equiv \delta \! * \! A $.
Then, we have the following equivalent conditions:
(i) $Z_1 \wedge Z_2 =0$,
(ii) A constant duality rotation $\theta$ exists such that $F = \cos \theta A + \sin \theta *A$ is a Killing-Yano tensor,
(iii) $K = F^2$ is a Killing tensor,
(iv) The Killing two-form $\dif Z$ of the Killing vector $Z = \delta * F$ is aligned with $F$: $[F, \dif Z] = 0$.
\\
Moreover, if these conditions hold, $Z$ and $Y=K(Z)$ are Killing vectors (or they vanish) such that $[Z,Y]=0$.
\end{proposition}
{\bf Remark 4.} Some of the results in propositions \ref{prop-inv} and \ref{prop-KNUT} were obtained in \cite{hougshton-sommers}
and \cite{collinson-smith} for the case of the ${\cal D}$-metrics. In \cite{fsEM-sym} we completed and partially extended these
results, and here we state that all of them hold for both, the $\tilde{\cal D}$-metrics and the conformally flat case.

In the present version of the generalized Jebsen-Birkhoff theorem the additional symmetry is defined by a hypersurface-orthogonal
 Killing vector. When do the Killing vectors in theorem \ref{theo-Killing} have this property? We know that a {\it simple}
 Killing-Yano tensor exists in the A-metrics and B-metrics where the Jebsen-Birkhoff theorem applies. Let us note that $A$ is
 a simple (rank two) two-form if, and only if, the scalar invariant
 $(A, *A) = \frac12 \eta_{\alpha \beta \gamma \delta} A^{\alpha \beta} A^{\gamma \delta}$ vanishes.
  We show now that this condition for the CKY tensor guarantees the hypersurface-orthogonal character of the
  Killing vectors defined by ${\cal Z} = \delta {\cal A}$.

It is worth remarking that $(A, *A)=0$ is equivalent to $\lambda$ being a real scalar and, from  (\ref{um-max}),
implies that the vector ${\cal U}(\delta {\cal U}) $ is real, that is, the 2+2 structure is integrable
(the two 2-planes are foliation) \cite{fs-KY}. On the other hand, $Z_1 = \delta A$ and $Z_2 = \delta \! *\! A$ are
 hypersurface-orthogonal vectors when $I_1 = *(Z_1 \wedge \dif Z_1) = 0$ and $I_2 = *( Z_2 \wedge \dif Z_2) = 0$.

By using (\ref{Z}) and (\ref{dZ}) we can compute the (real) vectors $I_1$ and $I_2$ and the codiferential of the complex
vectors ${\cal U}({\cal Z})$ and $\bar{\cal U}({\cal Z})$, and we obtain:
\begin{equation} \label{Is-2}
\begin{array}{l}
 \hspace*{-1.4mm}I_1 + I_2 =  2 \ci \{m\, \bar{\cal U}({\cal Z}) -a\,  {\cal U}({\cal Z})   + \frac23
e^{-\bar{\lambda}}[ ({\cal Z},{\cal Z}) \bar{\cal U}(\bar{\cal Z}) - ({\cal Z},\bar{\cal Z})  \bar{\cal U}({\cal Z})]  -   c. c. \} ,   \\[2mm]
\hspace*{-1.4mm}I_1 - I_2 = 2 \ci \{m \, \bar{\cal U}(\bar{\cal Z}) -a\,  {\cal U}(\bar{\cal Z}) + \frac23
 e^{-\bar{\lambda}}[ ({\cal Z},\bar{\cal Z}) \bar{\cal U}(\bar{\cal Z}) - (\bar{\cal Z},\bar{\cal Z})  \bar{\cal U}({\cal Z})]  -  c. c.\}  .
\end{array}
\end{equation}
\begin{equation} \label{dZU}
\delta \, {\cal U}({\cal Z}) = a + \frac23  e^{-\lambda} ({\cal Z},{\cal Z}) \, , \quad
\delta \, \bar{\cal U}({\cal Z}) = m + \frac23  e^{- \bar{\lambda}} ({\cal Z}, \bar{\cal Z})   .
\end{equation}

Let us suppose that $A$ is a simple CKY tensor and $[ {\cal A} , Ric]=0$. From theorem \ref{theo-Killing}
this second condition implies that $Z_1$ and $ Z_2$ are Killing vectors. Moreover, as commented above, the first
 one implies that $\lambda$ and ${\cal U}(\delta {\cal U}) $ are real, and (\ref{um-max}) and (\ref{Z}) impose ${\cal U}({\cal Z})$
 and $\bar{\cal U}({\cal Z})$ are real too. Then (\ref{dZU}) imposes $a$ and $m$ to be real scalars. Using all these
 constraints in (\ref{Is-2}) we obtain $I_1 = I_2 = 0$ and consequently $Z_1$ and $Z_2$ are hypersurface-orthogonal vectors.

Let us supose now that $Z_1, Z_2$ are hypersurface-orthogonal Killing vectors. Thus $I_1 = I_2 = 0$. Moreover,
 from theorem \ref{theo-Killing} we have $[ {\cal A} , Ric]=0$. Then $Z_1$ and $Z_2$ lie on the principal planes
 defined by the Ricci and/or the Weyl tensors (which are that defined by ${\cal U}$). This condition
 imposes ${\cal U}({\cal Z})$ and $\bar{\cal U}({\cal Z})$ to be real when $Z_1 \wedge Z_2 \not=0$, and then
 from (\ref{Is-2}) and (\ref{dZU}) we can show that $\lambda$ is real and then $(A, *A)=0$. When $Z_1 \wedge Z_2 =0$, we
 can consider the Killing-Yano tensor $F$ and $Z \equiv \delta \! * \! F $ of proposition \ref{prop-KNUT}, and then a similar reasoning implies $(F, *F)=0$ if $(Z,Z)\not=0$. Moreover $(Z, Z) =0$ implies ${\cal U}(Z) = Z$,  and then (\ref{dZU}) leads to $\nabla Z =0$, that is, the space-time is a pp-wave if $Z \not= 0$. Otherwise, when $Z=0$,  $*F$ is also a Killing-Yano tensor and the space-time is a product one. Consequently we have proven the following.

\begin{theorem} \label{theo-Killing-integrable}
In a non constant curvature space-time admitting a non null conformal Killing-Yano tensor $A$ such that $[A , Ric]=[*A, Ric]=0$,
 let us consider $Z_1 \equiv \delta A$ and $Z_2 \equiv \delta \! * \! A $. Then, the following statements hold:

If $Z_1 \wedge Z_2 \not=0$, then $Z_1$ and $Z_2$ are hypersurface-orthogonal Killing vectors if, and only if, $(A, *A)=0$.

If $Z_1 \wedge Z_2 =0$, let $F = \cos \theta A + \sin \theta *A$ be the Killing-Yano tensor given in proposition \ref{prop-KNUT},
and $Z \equiv \delta \! * \! F \not=0$. Then: (i) when $(Z,Z) \not=0$,  $Z$ is a hypersurface-orthogonal Killing vector if, and only if, $(F, *F)=0$; (ii) when $(Z,Z) =0$,  $Z$ is a hypersurface-orthogonal Killing vector and the space-time is a pp-wave.

The space-time has a product metric if, and only if, $Z_1=Z_2=0$. 
\end{theorem}
%
%
%
{\bf Remark 5.} When $Z_1 \wedge Z_2 \not=0$, the hypersurface-orthogonal nature of the Killing vectors $Z_1$ and $Z_2$ is
 guaranteed if $(A, *A)=0$, that is, when the scalar $\lambda$ is real. Nevertheless, if the imaginary part of $\lambda$ is
 a non vanishing constant (and then $(A, *A)\not=0$), a constant duality rotation leads to a simple CKY tensor
 $A' = \cos \theta A + \sin \theta \! * \! A$, and the first statement in theorem \ref{theo-Killing-integrable} applies.
\
\\[2mm]
{\bf Remark 6.} When a Killing tensor $F$ exists and $Z \equiv \delta \! * \! F \not=0$ is a null Killing vector ($(Z,Z) =0$), then $\nabla Z = 0$ and the space-time is a pp-wave. In this case $F$ has a constant eigenvalue and $F$ is simple only when this eigenvalue vanishes. The only Ricci tensor of the considered type $[(11)(11)]$ that is compatible with the integrability conditions of $\nabla Z = 0$ has a vanishing eigen-value associated with the time-like eigen-plane. Consequently, these metrics have an unclear physical meaning.  
\
\\[2mm]
{\bf Remark 7.} Theorems \ref{theo-Killing} and \ref{theo-Killing-integrable} imply the existence of one or two symmetries
 if the space-time has some specific geometric properties. But these symmetries do not exist when both, $Z_1 = \delta A$ and
 $Z_2 = \delta \! *\! A$, vanish. In this case both, $A$ and $*A$, are Killing-Yano tensors and, as stated in the third point
  of theorem \ref{theo-Killing-integrable}, the space-time metric is a product one. It is worth remarking that this is, precisely,
   the case where the generalized Jebsen-Birkhoff theorem does not apply (see \cite{barnes} and \cite{bona}).

On the other hand, a simple Killing-Yano tensor $A$ exists in the space-times where the Jebsen-Birkhoff theorem applies. Moreover,
for a Killing-Yano tensor we have \cite{stephani} $[A, Ric]=0$. Then we obtain the following.

\begin{corollary} \label{cor-Birkhoff}
If a non constant curvature space-time admits a non null Killing-Yano tensor $A$ and $Z \equiv \delta\! * \! A \not=0$, then:

When $(Z,Z)\not=0$, then $Z$ is a hypersurface orthogonal Killing vector if, and only if, $[*A , Ric]=0$ and $(A, *A)=0$.

When $(Z,Z)=0$, then $Z$ is a hypersurface orthogonal Killing vector if, and only if, $[*A , Ric]=0$.
\end{corollary}

Plaintly, the generlized Birkhoff theorem follows from this corollary. Ideed, a metric $g$ admitting a maximal group of symmetries on two-dimensional non-null orbits admits the canonical form of a 2+2 warped product \cite{bona}:
\begin{equation} \label{metric}
g = v(x^0, x^1) + \phi^2(x^0, x^1) h(x^2,x^3) \, ,
\end{equation}
where $\phi$ is a function, $v$ is a Lorentzian (or Riemannian) metric and $h$ is a Riemannian (or Lorentzian) metric of constant curvature. Then  $A=\phi H$ is a simple Killing-Yano tensor for the metric (\ref{metric}), $H$ being the metric volume element of the metric $\phi h$ \cite{fs2+2}. On the other hand, the Ricci tensor of the metric (\ref{metric}) has $\{x^2,x^3\}$ as an eigenplane. Consequently, a Ricci with two double eigen-values implies its alignement with $*A$. Moreover, we have $Z \equiv \delta\! * \! A = -*(\dif \phi \wedge H)$, which vanishes if, and only if, $\dif \phi=0$. Thus, we can apply corollary \ref{cor-Birkhoff} and we recover the generalized Birkhoff theorem. 

\begin{corollary} \label{cor-Birkhoff-2}
If a space-time admits a maximal group of symmetries acting on two-dimensional non-null orbits and the Ricci tensor is of types $[(11)(11)]$ or $[1111]$ then it admits an additional hipersurface-orthogonal Killing vector provided that $\dif \phi \not=0$. This Killing vector is given by $Z=-*(\dif \phi \wedge H)$, where $H$ is the volume element of the group orbits.
\end{corollary}

\section{Ending comments.}

Corollary generalizes the Jebsen-Birkhoff theorem because the hypothesis of a three-dimensional group
 of isometries on two-dimensional orbits has been weakened by considering the existence of a simple Killing Yano tensor. Note
 that no symmetries are required a priori. Nevertheless, if we consider the non-conformally flat case with $\Lambda = constant$,
 we obtain the charged A-metrics and B-metrics, and all of them admit maximal symmetry on non null two-dimensional orbits. But
 these symmetries are not a hypothesis but a consequence of the field equations.

Corollary \ref{cor-Birkhoff} not only generalizes the Jebsen-Birkhoff theorem by weakening its hypothesis, but it also offers
two other new improvements. On the one hand, it is stated as a necessary and sufficient condition. On the other hand, it gives the explicit expression of the hypersurface-orthogonal Killing vector in
terms of the magnitudes which appear in the hypothesis (the simple Killing tensor). This advance can be also found in the statement of the Jebsen-Birkhoff theorem given in corollary \ref{cor-Birkhoff-2}.  

Moreover, theorem \ref{theo-Killing-integrable} is a wider generalization than corollary \ref{cor-Birkhoff} because it establishes the existence of
one or two hypersurface-orthogonal Killing vectors under weaker conditions. Besides, theorem \ref{theo-Killing} includes theorem
\ref{theo-Killing-integrable} as a particular case and it shows the close relationship between the Jebsen-Birkhoff theorem and
the results by Tachibana \cite{tachibana}  and Hougshton and Sommers \cite{hougshton-sommers}.

It is worth remarking that the different cases we found in our study correspond with extensions of the invariant classes of
vacuum type D solutions presented in \cite{fs-buit}. For each of these classes there is the charged counterpart in the set
of the ${\cal D}$-metrics, and with similar invariant definitions we can consider the same classes in the $\tilde{\cal D}$-metrics
and in the conformally flat case. Thus, when $Z_1 \wedge Z_2 \not=0$ we have C-like metrics admitting the subclass where $Z_1$ and
$Z_2$ are hypersurface-orthogonal vectors (containing the strict Ehlers and Kundt C-metrics). When $Z_1 \wedge Z_2 =0$ we have
Kerr-NUT-like metrics, with a regular case characterized by $Z \wedge K(Z) \not = 0$. Note that in these three quoted cases a
two-dimensional commutative group of isometries exists (see propositions \ref{prop-inv} and \ref{prop-KNUT}). However,
 when $Z_1 \wedge Z_2 = Z \wedge K(Z) = 0$ we obtain A-NUT-like and B-NUT-like metrics and only a real Killing vector $Z$ is
 defined by the CKY tensor $A$. When $Z$ is a hypersurface-orthogonal vector we have the subfamily where corollary \ref{cor-Birkhoff} applies.

Theorem \ref{theo-Killing} also holds when $A$ is a null CKY tensor. Nevertheless, the other resuts presented herein only apply
when $A$ is a non null CKY tensor, and their possible generalization to the null case is not evident. This study and its relationship
with previously known Birkhoff-like results for null orbits \cite{barnes} \cite{barnes2} will be considered elsewhere.

\begin{acknowledgements}
 This work has been supported by the Spanish ``Ministerio de
Econom\'{\i}a y Competitividad", MICINN-FEDER project FIS2012-33582.
\end{acknowledgements}


\end{document}